\pdfoutput=1
\documentclass[a4paper,10pt]{article}
\usepackage[T1]{fontenc}
\usepackage[utf8]{inputenc} 
\usepackage{lmodern}
\usepackage[english]{babel}
\usepackage{graphicx}
\usepackage{amsmath}
\usepackage{float}
\usepackage{authblk}
\usepackage{tabularx}
\usepackage[backend=biber,sortcase=true,sortcites=true,sorting=none,style=science,dateabbrev=false]{biblatex}
\usepackage{times}
\usepackage{bm}
\usepackage{xcolor}
\usepackage{csquotes}
\usepackage{epstopdf}
\usepackage[final]{pdfpages}

\providecommand{\keywords}[1]
{
  \small	
  \textbf{\textit{Keywords---}} #1
}

\author[1,2]{Alejandro Alés}
\author[1,2]{Fernando Lanzini}
\affil[1]{Instituto de Física de Materiales Tandil (IFIMAT), Universidad Nacional del Centro de la Provincia de Buenos Aires (UNCPBA), Pinto 399, 7000 Tandil, Argentina}
\affil[2]{Consejo Nacional de Investigaciones Científicas y Técnicas (CONICET), Argentina}
\date{}                     
\bibliography{ms.bib}

\DeclareUnicodeCharacter{0301}{'}

\DeclareUnicodeCharacter{E5F8}{ }

\begin{document}

\title{Mechanical and thermodynamical properties of $\beta-Cu-Al-Mn$ alloys along the $Cu_3Al \to Cu_2AlMn$ compositional line}
  

    \maketitle
\begin{abstract}
    The elastic properties of $Cu-Al-Mn$ alloys with compositions along the  $Cu_3Al \to Cu_2AlMn$ line and $bcc$-based structures, are studied by means of first-principles calculations. From the calculated elastic constants, the Zener's anisotropy, sound velocities and Debye temperature are determined.  The theoretical results compare well with the available experimental data. The influence of vibrations is introduced through the quasi-harmonic Debye model, and different properties are studied as functions of temperature and composition.
\end{abstract}

\keywords{Cu-Al-Mn, First-Principles Calculations, Elastic Constants, Debye Temperature, Quasi-Harmonic Approximation}
\\

\section{Introduction}
\label{sc:Int} 

\indent Shape memory alloys (SMAs) have attracted attention in the last decades due to their interesting mechanical properties (pseudoelasticity, shape memory effect, double shape memory effect) \cite{jani2014review}. These mechanical properties are associated with a difussionless martensitic transformation \cite{delaey1991materials}. $Cu$-based alloys are particularly interesting due to their lower cost and comparatively good shape memory properties \cite{sutou2008ductile}. In these systems, the martensitic transformation takes place from a metastably retained $\beta$ phase, with bcc structure, to a martensitic phase with close-packed structure.  One of such Cu-based SMAs is the $Cu-Al-Mn$ system: the use of this family of alloys in seismic applications \cite{araki2011potential,hosseini2015experimental,liu2015superelastic,aslani2012potential} as well as in medicine, aeronautics and robotics \cite{shrestha2012application,araki2014feasibility,chang2016use,oliveira2016laser,pareek2018plastic} has been subject of extensive research in the last years. \\
\indent The $Cu-Al-Mn$ system presents some advantageous properties as compared to other $Cu$-based shape memory alloys. The addition of $Mn$ increases the range of stability of the $\beta$ phase. Besides, the presence of $Mn$ confers magnetic properties, as a result of the coupling between the magnetic moments located at the $Mn$ sites. For instance, the Heusler alloy with stoichiometric composition $Cu_2AlMn$  has a ferromagnetic to paramagnetic transition with a high Curie temperature $T_C \approx 630 K$ \cite{konoplyuk2011magnetoresistance}. Other interesting feature is the formation of a miscibility gap at temperatures around $600 K$ and below, for compositions along the pseudobinary line $Cu_3Al \to Cu_2AlMn$ \cite{bouchard1972interphase,obrado1999quenching,sato1994quantification,marcos2004kinetics,velazquez2017spinodal}. This two-phases gap is formed by a spinodal decomposition mechanism, giving place to regions with composition close to $Cu_3Al$ and $DO_3$ structure, and others with composition near $Cu_2AlMn$ and $L2_1$ order \cite{oxley1963heusler,kainuma1998phase,kudryavtsev2005effect,dubois1979decomposition,velazquez2017spinodal,velazquez2020spinodal}. The $Cu_3Al$ regions are paramagnetic, whereas the $L2_1$ ones are ferromagnetic. Although the nature of the gap is still a matter of discussion, it is known that the difference in lattice parameters between the phases leads to internal tensions that could partly explain the decomposition \cite{bouchard1972interphase,kainuma1998phase,velazquez2017spinodal}. In a previous work \cite{lanzini2015role} it has been shown that there is also magnetic reasons behind this decomposition.       \\
\indent The aim of this work is to characterize, using first-principles methods and the quasiharmonic approximation (QHA), the elastic, vibrational and thermodynamic properties of alloys along the $Cu_3Al \to Cu_2AlMn$ line of compositions. This will allow to gain insight about the reasons leading to the formation of the miscibility gap.\\ 
 \indent The rest of this work is organized as follows: in section \ref{sc:Met} the computational details of the first principles calculations are explained, and the main equations of the QHA are introduced. In section \ref{sc:RC} the results are presented and discussed, and the main conclusions are outlined in section \ref{sc:Concl}.

\section{Methodology}
\label{sc:Met} 
\subsection{First-principles calculations}

\begin{figure}[h!]
\centering
\includegraphics[width=0.45\textwidth]{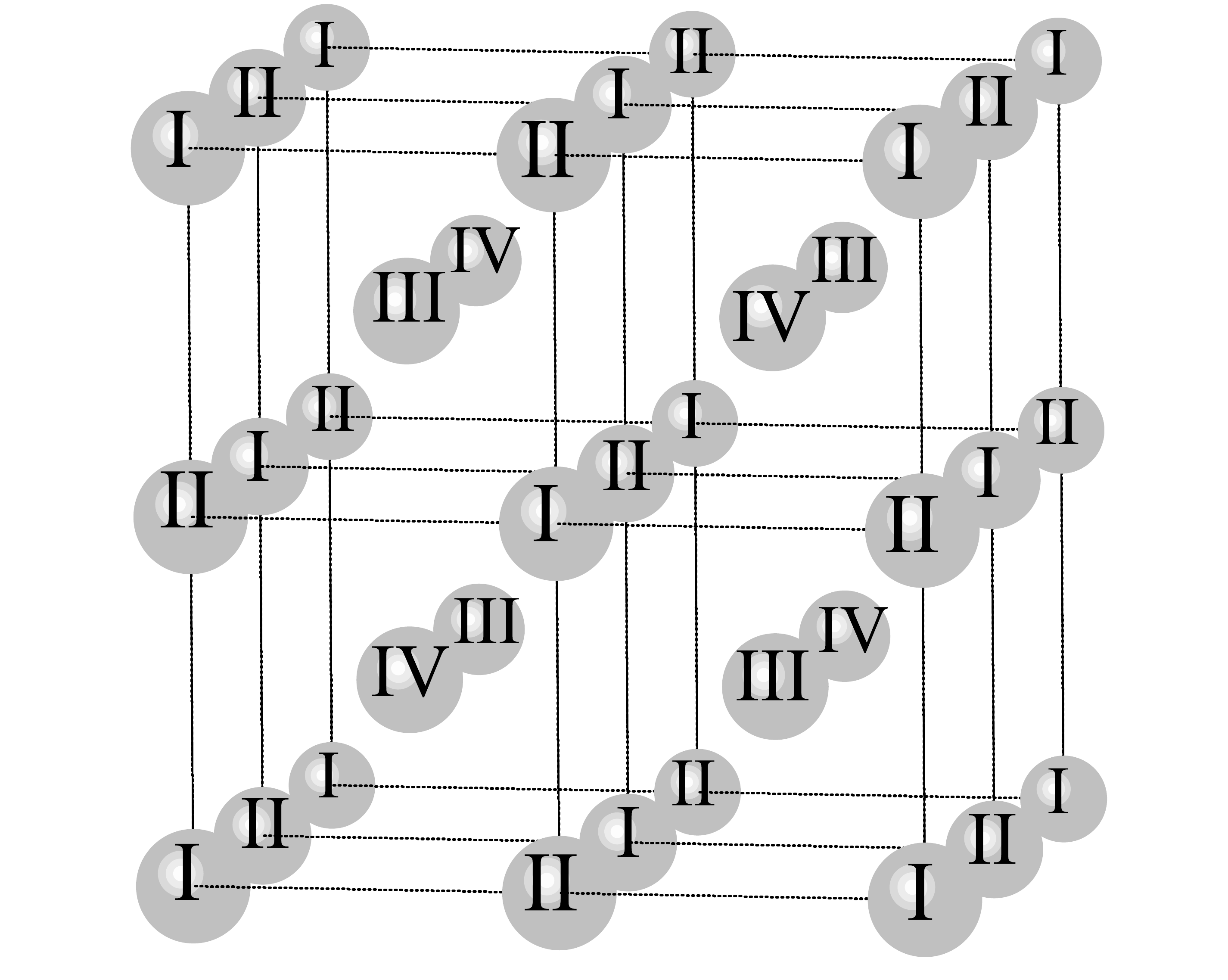}
\caption{Body centered cubic supercell subdivided in four interpenetrating fcc sublattices}
\label{fig:Supercell}
\end{figure}

\indent First-principles calculations were performed by means of the Quantum Espresso implementation \cite{qe1,qe2}. This is an integrated suite of computational codes based on the density functional theory (DFT) \cite{PhysRev.136.B864, PhysRev.140.A1133}, and employs an expansion of plane waves and pseudopotentials. In the present work, the ionic cores for $Cu$, $Al$ and $Mn$ were described by Vanderbilt ultrasoft pseudopotentials \cite{vanderbilt1990soft}. The exchange-correlation term of the spin-polarized calculations was represented by the Perdew-Burke-Ernzerhorf implementation of the generalized gradient approximation (GGA) \cite{PhysRevLett.77.3865}. A careful examination of the energy convergence respect to different control parameters was performed prior to the self-consistent calculations. The energy cut-off for the plane wave expansion was established in $40$ Ry, and for the charge density in $480$ Ry; a uniform mesh of $10 x 10 x 10$ k points, automatically generated according to a Monkhorst-Pack scheme was employed \cite{monkhorst1976special}. The convergence criteria in the total energy for the self-consistent cycle was set to $1 x 10^{-8} Ry$.\\
\indent As postulated by other authors \cite{deb2000, zukowski1997} and verified in a previous work \cite{lanzini2015role}, the magnetic contributions from $Cu$ and $Al$ atoms are negligible; then, the spin-polarized calculations were  performed assuming that the magnetic moments  are due solely to $Mn$ atoms.\\
\indent Five alloys along the line of compositions ${Cu_{3}Al \to Cu_{2}AlMn}$ were studied. Description of these structures can be made with the help of Figure \ref{fig:Supercell}. In the limiting $Cu_{3}Al$ alloy with $DO_{3}$ order, sublattices $I$, $II$ and $III$ are occupied by $Cu$ atoms and sublattice $IV$ by $Al$. In the Heusler $Cu_{2}AlMn$ alloy with $L2_{1}$ order, sublattices $I$ and $II$ are occupied by $Cu$, sublattice $III$ by $Mn$, and sublattice $IV$ by $Al$. Besides these limiting cases, three ordered compounds with compositions $Cu_{11}Al_{4}Mn_{1}$, $Cu_{10}Al_{4}Mn_{2}$ and $Cu_{9}Al_{4}Mn_{3}$ were also studied (note that all the considered alloys posses a fixed aluminum content of 25 at. \%). The three intermediate systems, with 16 atoms per unit cell, were constructed starting from $Cu_{3}Al$ and replacing, respectively, 1, 2 or 3 $Cu$ atoms in the sublattice $III$ (Fig. 1) by $Mn$ atoms. The corresponding compositions lie equidistantly between $Cu_{3}Al$ and $Cu_{2}AlMn$. The resulting structures can be regarded as partially ordered $L2_{1}$ structures.\\
\indent For each of the studied alloys, a structural optimization varying the cubic lattice parameter was done. The equilibrium state was determined by locating the minimum of the energy as a function of the lattice parameter, fitting the calculated data with a Murnaghan equation of state (EOS) \cite{Murnaghan244}. From these fittings, the energy per formula unit $E_{0}$, the equilibrium volume $V_0$, the equilibrium bulk modulus $B_0$ and its pressure derivative $B_0^\prime$, were obtained.

\subsection{Determination of the elastic constants}
\label{sb:CE}

\indent  In a cubic system there are only three independent elastic constants, namely $C_{11}$, $C_{12}$ and $C_{44}$. Determination of the elastic constants is performed by applying different distortions to the cubic cell. The bulk modulus $B_0$ and the equilibrium volume $V_{0}$ are obtained by fitting the Murnaghan EOS \cite{Murnaghan244}, and correspond to an hydrostatic variation of volume by means the following perturbation of the basis vectors of the lattice

 \begin{center}
$D(\delta) $=$
\begin{pmatrix}
   1+\delta  & 0 & 0 \\
  0 &  1+\delta  & 0 \\
  0 & 0 &  1+\delta  
 \end{pmatrix} $
  \end{center}

\indent  where $\delta$ is the magnitude of the strain. $B_{0}$ is related with the elastic constants by the expression  

\begin{equation}
B_0 = \frac{C_{11} + 2 C_{12}}{3} 
\label{Bulk}
\end{equation}

\indent  The combination $C_{11}-C_{12}$ is obtained by performing a volume-conserving orthorhombic strain \cite{mehl1990structural,rached2009first}
 \begin{center}
$D(\delta) $=$
\begin{pmatrix}
   1+\delta  & 0 & 0 \\
  0 &  1-\delta  & 0 \\
  0 & 0 &  \frac{1}{1-\delta^2}  
 \end{pmatrix} $
  \end{center}
  
\indent    and fitting to the change in energy with respect to the unstrained crystal
  
  \begin{equation}
   \Delta E(\delta) = \Delta E (-\delta) = V_0 (C_{11} - C_{12} ) \delta^2 + O(\delta^4) 
  \label{DistOrto}
\end{equation}

\indent  The combination of  Eqs. \eqref{Bulk} and \eqref{DistOrto} allows calculating $C_{11}$ and $C_{12}$. The remaining elastic constant, $C_{44}$, is obtained by applying a volume-conserving monoclinic distortion

 \begin{center}
 $D(\delta) $=$
\begin{pmatrix}
   1  & \frac{\delta}{2} & 0 \\
  \frac{\delta}{2} &  1  & 0 \\
  0 & 0 &  \frac{4}{4-\delta^2}  
 \end{pmatrix} $
 \end{center}

\indent  In this case, the difference in energy respect to the relaxed lattice is expressed as

\begin{equation}
    \Delta E (\delta) = \Delta E (-\delta)= \frac{1}{2} V_0 C_{44} \delta^2  + O(\delta^4)
    \label{DistMono}
\end{equation}

\indent  In the present work, ten distortions of each type (orthorhombic or monoclinic), with values $\pm 1 \%, \pm 2 \%, \pm 3 \%, \pm 4 \%$ and $ \pm 5 \%$, were applied to the relaxed cubic structure with volume $V_{0}$.\\ 
\indent  From the values of the elastic constants the shear modulus is obtained. We will use the Hill method \cite{hill1963elastic}, according to which the shear modulus $G_{H}$ is given by the average 
\begin{equation}
G_{H} = \frac{G_{v}+G_{r}}{2}, \label{GH}
\end{equation}

\indent  where $G_{v}$ is the Voigt's estimate \cite{voigt1928lehrbuch}:
\begin{equation}
    G_v = \frac{C_{11} - C_{12} + 3 C_{44}}{5}
    \label{GV}
\end{equation}

\indent  and $G_{r}$ is the Reuss value \cite{reuss1929berechnung}
\begin{equation}
    G_r = \frac{5 \left( C_{11}-C_{12}\right) C_{44}}{3 \left( C_{11}-C_{12}\right)+ 4 C_{44}}
    \label{GR}
\end{equation}

\indent  The isotropic Young modulus, $E$, and Poisson ratio, $\nu$, are given by 

\begin{equation}
     E = \frac{9 B_{0} G_{H}}{3 B_{0} + G_{H}}
    \label{Young}
\end{equation}

\begin{equation}
     \nu = \frac{3 B_{0} - 2 G_H}{2 (3 B_{0}+G_{H})}
    \label{nuIsot}
\end{equation}

\subsection{Thermal properties}
\label{sb:Thermalproperties} 

\indent  The vibrational motion of the lattice was incorporated through the quasi-harmonic Debye model, as implemented in the software GIBBS \cite{blanco2004gibbs}. Taking as input the ab-initio values of $E(V)$ for the hydrostatic variation of volume and the Poisson ratio $\nu$, Eq. \eqref{nuIsot}, the adiabatic bulk modulus is computed by means of the derivative
 
\begin{equation}
 B_{S} \approx  V \frac{d^{2}E(V)} {dV^{2}}
\end{equation}

\indent  and then the volume-dependent Debye temperature is obtained 

\begin{equation}
 \Theta_D = \frac{\hbar}{k} \left[ 6 \pi^2 V^{1/2} n \right]^{1/3} \left( \frac{B_{S}}{M} \right)^{1/2} f(\nu) 
 \label{eq:Recio}
\end{equation}

\indent  where 

\begin{align*}
& f(\nu) = \\ &\left\{ 3 \left[2 \left( \left( \frac{2}{3} \right) \frac{1+\nu}{1-2\nu} \right)^{3/2} + \left( \left( \frac{1}{3} \right) \frac{1+\nu}{1-\nu} \right)^{3/2} \right]^{-1} \right\}^{1/3}
\end{align*} 

\indent  The thermal evolution of different properties can be evaluated by constructing the Gibbs free energy as 

\begin{equation} 
G(V,P,T)=E(V) + PV + A_{vib}(\theta_{D}(V); T)
\label{gibbs}
\end{equation}

\indent  being $A_{vib}$ the vibrational contribution

\begin{align}
A_{vib}& =  n k_B T \notag\\ & \left[ \frac{9}{8}  \frac{\Theta_D}{T} - 3 log\left(1-e^{-\frac{\Theta_D}{T}}   \right) - D\left(\frac{\Theta_D}{T} \right)\right]    
\end{align}

\indent  \indent  with $ D\left(\frac{\Theta_D}{T}\right)$ being the third-order Debye integral, $n$ the number of atoms in the unitary cell and $k_B$ the Boltzmann's constant. At given $(p, T)$ the equilibrium volume is found from the condition 

\begin{equation}
    \left( \frac{\partial G(V,P,T)}{\partial V} \right)_{p,T} = 0
\end{equation}

\indent  After that, it is possible to calculate other properties such as the isothermal bulk modulus

\begin{equation}
    B_{T} (p,T) = - V     \left( \frac{\partial^2 G(V,P,T)}{\partial V^2} \right)_{p,T}
\end{equation}

\indent  the vibrational heat capacity

\begin{equation}
C_v = 3 n k_B \left[ 4 D\left(\frac{\Theta_D}{T} \right) - \frac{3\frac{\Theta_D}{T} }{e^{\frac{\Theta_D}{T}} -1}  \right]
\end{equation}

\indent  and the volumetric thermal expansion coefficient 

\begin{equation}
\alpha = \frac{\gamma C_v}{B_T V}
\end{equation}

\indent  where $\gamma$ is the Grüneisen parameter, obtained through the Mie-Grüneisen equation \cite{francisco1998quantum, blanco2004gibbs}.

\section{Results and discussion}
\label{sc:RC} 

\subsection{Elastic and acoustic properties}

\begin{figure}[h!]
\centering
\includegraphics[width=0.45\textwidth]{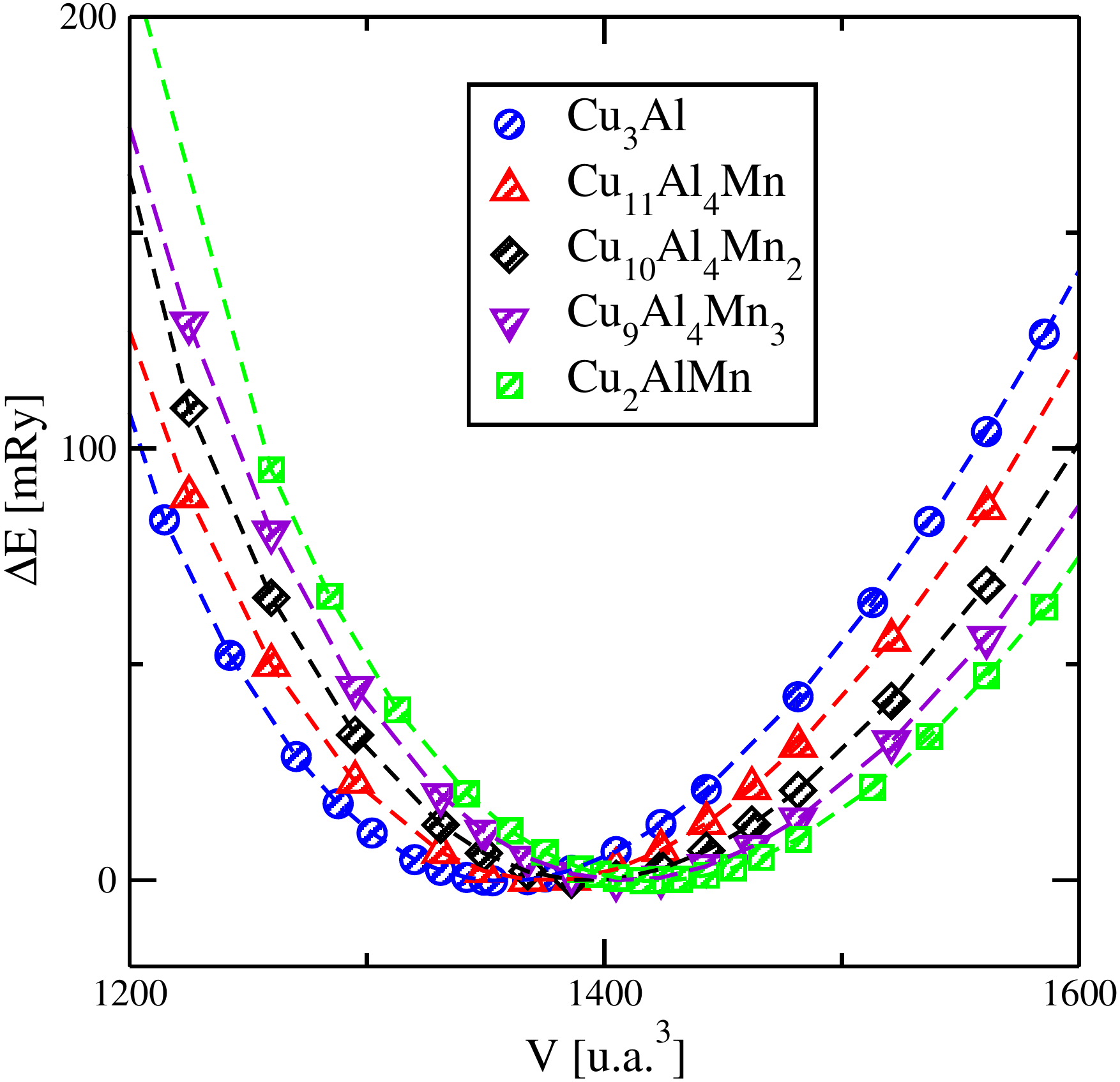}
\caption{Energy difference from the equilibrium volume as function of the volume for different compositions. The dashed line corresponds to the Murnaghan fit.}
\label{fig:Bulk}
\end{figure}

\begin{figure}[h!]
\centering
\includegraphics[width=0.45\textwidth]{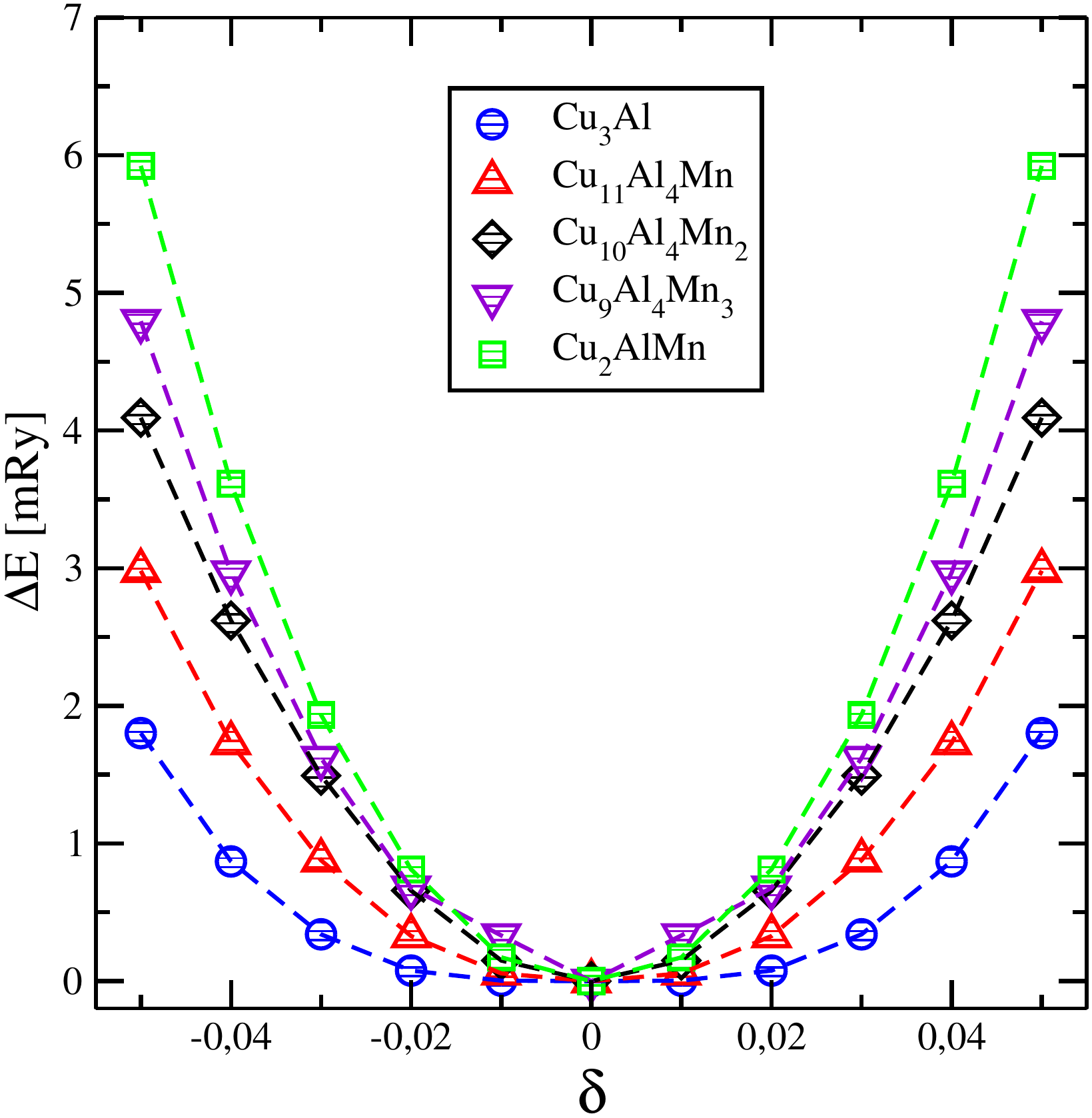}
\caption{Energy differences for orthorhombic deformations $\delta$ and the different compositions. The dashed lines are just guides for eyes.}
\label{fig:Ort}
\end{figure}

\begin{figure}[h!]
\centering
\includegraphics[width=0.45\textwidth]{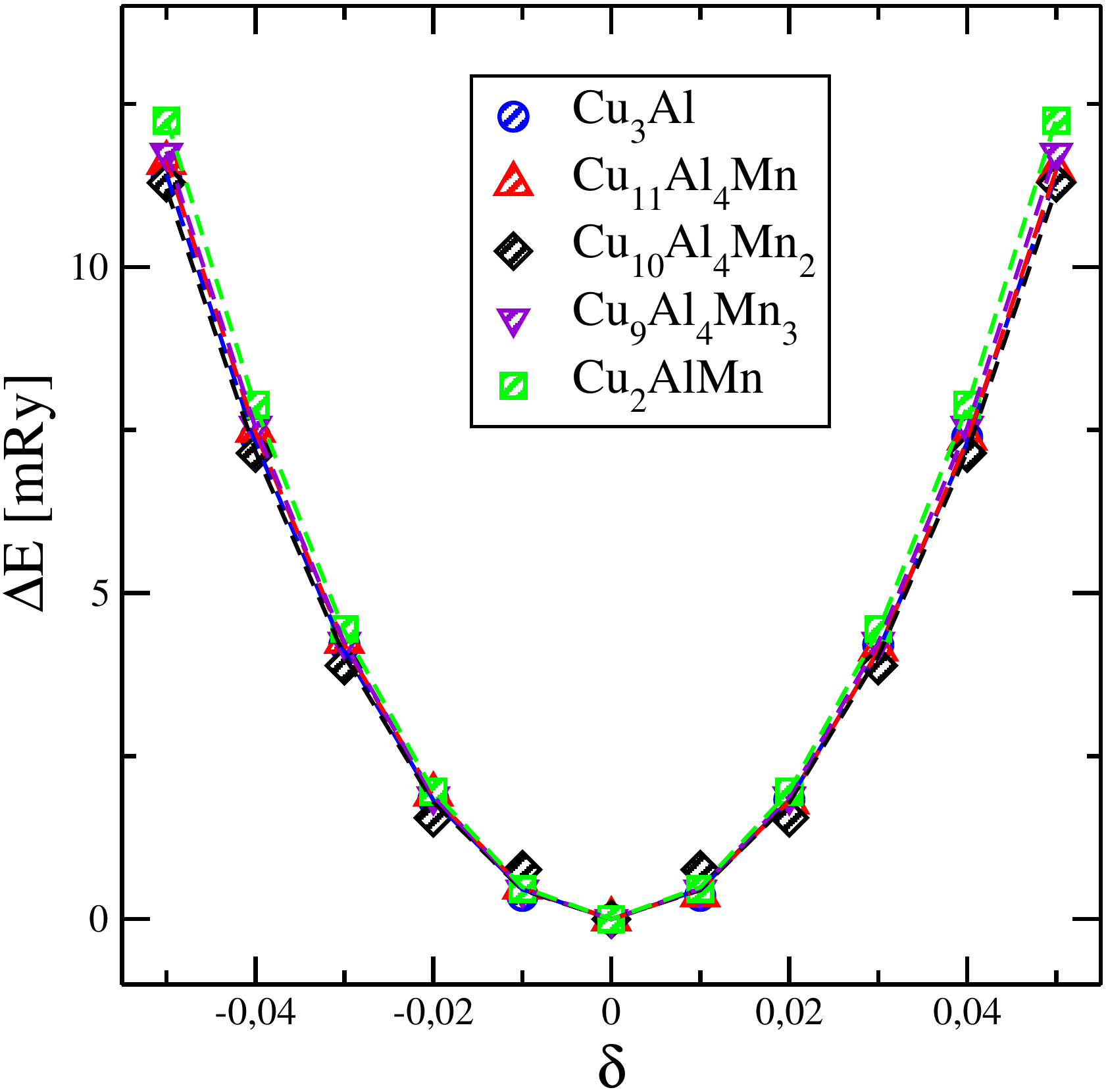}
\caption{Energy difference for monoclinic deformations $\delta$ and for the different compositions. The dashed lines are just guides for eyes.}
\label{fig:Mon}
\end{figure}

\indent  In Figures \ref{fig:Bulk} - \ref{fig:Mon}, the changes in energy in reference to the equilibrium volume for different strains and alloy compositions are shown. The hydrostatic changes of volume shown in Fig. \ref{fig:Bulk} were fitted by the Murnaghan EOS for each alloy, determining the corresponding values of $V_{0}$ and $B_{0}$. These values are listed in Table \ref{tab:C-E}. When orthorombic strains are applied (Fig. \ref{fig:Ort}), the variation of energy with $\delta$ is more markedly composition dependent than for the monoclinic strains (Fig. \ref{fig:Mon}), where the curves for different compositions almost overlap. This implies that the $C_{11}-C_{12}$ difference varies more with composition than the $C_{44}$ value. Furthermore, the difference $ C_{11} -C_{12} $ has a monotonous behavior respect the copper contents.

\indent  The values of the elastic constants obtained through Eqs. \ref{Bulk} - \ref{DistMono} are listed in Table \ref{tab:C-E}. Experimental values and theoretical calculations taken from the literature are included for comparison. For the Heusler alloy $Cu_{2}AlMn$, our results agree well with the experimental determinations by Michelutti et al \cite{michelutti1978magnetization} (values extrapoled to $0 K$), as well as with recent theoretical calculations using full-potential linearized augmented plane waves (FPLAPW) \cite{wu2019elastic, jalilian2015comment} and the projector augmented wave method (PAW) \cite{wen2018first}. For the intermediate compositions, the elastic constants have been experimentally determined for alloys $Cu-25 \% at Al-5 at. \% Mn$ (at temperatures above the martensitic transformation) and for $Cu - 25 at. \%Al-7.5 at. \% Mn$ and $Cu - 25 at. \%Al-10 at. \% Mn$ (above liquid nitrogen) \cite{prasetyo1976elastic}. Extrapolation of these experimental results to $0 K$ are also listed in Table \ref{tab:C-E}.   

\begin{table*}
\begin{center}
\begin{tabular}{ | c | c | c | c | c | c | c | c |}\hline Alloy & Ref. & $V_0 [ua^{3}] $ & $ B_0 [GPa] $ & $ C_{11}[GPa]$ & $ C_{12} [GPa] $ & $C_{44}[GPa]$   \\ \hline
$ Cu_3 Al $ & This work & $ 84.84348 $ &$ 131.23  $ & $ 137.23  $ & $ 128.03   $ & $ 99.08  $ \\\hline
$ Cu_{70} Al_{25} Mn_{5}  $ & \cite{prasetyo1976elastic}, experimental & $ --- $ & $ 129.33  $ & $ 140.0 $ & $ 124.0$  &  $ 98 $ \\ \hline
$ Cu_{11}Al_4 Mn $ & This work & $ 85.9660 $ & $ 128.11 $ & $135.97 $ & $ 124.35$  & $ 99.03 $  \\ \hline
$ Cu_{67.5} Al_{25} Mn_{7.5} $ & \cite{prasetyo1976elastic}, experimental & $ --- $ & $ 128.0  $ & $ 138.0 $ & $ 123.0$  &  $ 104 $ \\ \hline
$ Cu_{10}Al_4 Mn_2 $ & This work & $ 87.1304$  & $ 123.54 $ & $ 134.73 $ & $ 117.94 $  & $ 94.59 $ \\ \hline
$ Cu_{65} Al_{25} Mn_{10} $ & \cite{prasetyo1976elastic}, experimental & $ --- $ & $ 128.0  $ & $ 138.0 $ & $ 123.0$  &  $ 104 $ \\ \hline
$ Cu_9 Al_4 Mn_3 $ & This work & $ 88.0784 $ & $ 121.67 $ & $ 134.12 $ & $ 115.44 $ & $ 98.34 $  \\ \hline
& This work & $ 88.8360 $ & $ 120.51  $ & $ 138.8  $ & $ 111.3 $  &  $ 102.0  $ \\ 
& \cite{michelutti1978magnetization}, experimental \footnote{In the Ref. \cite{michelutti1978magnetization}, we use the extrapolated values from Ref. \cite{wu2019elastic}.} & $ 89.4498 $ & $ 110.4  $ & $ 128.1 $ & $ 101.5$  &  $ 104.4 $ \\ 
$ Cu_2 Al Mn $ & \cite{wu2019elastic}, FPLAPW & $ 88.1167 $ & $ 125.3  $ & $ 143.7 $ & $ 116.1$  &  $ 117.6 $ \\ 
&\cite{jalilian2015comment}, FPLAPW & $ --- $ & $ 122.3  $ & $ 137.0 $ & $ 115.0$  &  $ 112 $ \\ 
&\cite{wen2018first}, PAW & $ 88.44086 $ & $ 122.2  $ & $ 138.8 $ & $ 113.9$  &  $ 103.0 $ \\ \hline
\end{tabular}
\caption{Equilibrium volume per atom, bulk modulus, and elastic constants for the different alloys.}
\label{tab:C-E}
\end{center}
\end{table*}

\begin{figure}[ht]
\centering
\includegraphics[width=0.45\textwidth]{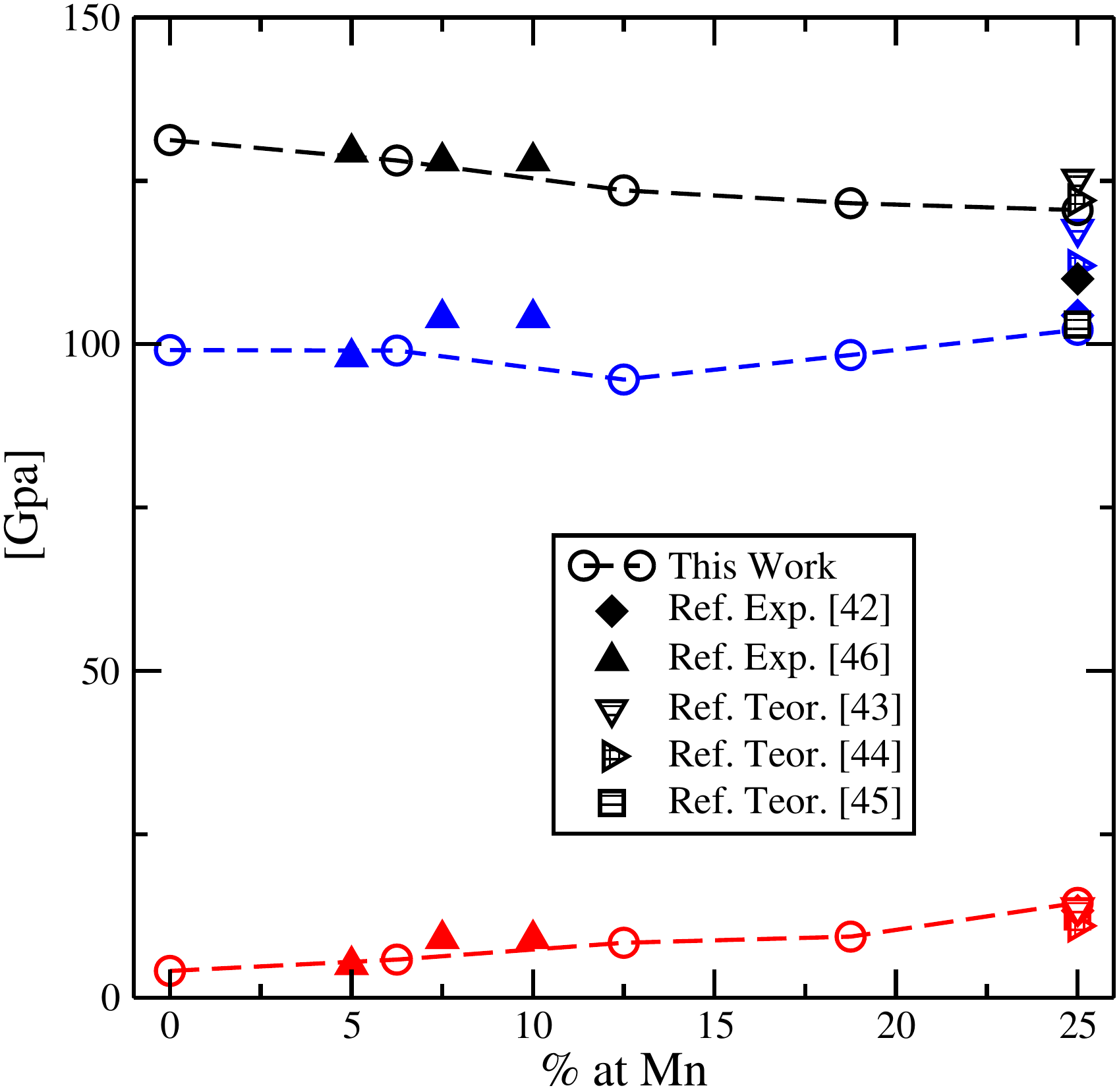}
\caption{Elastic moduli as a function of $Mn$ content. Code color: Black for Bulk modulus, Blue for $C_{44}$ and red for $C^\prime$. In empty circles, the present work results.}
\label{fig:ElasticModulii}
\end{figure}

\indent  In Fig \ref{fig:ElasticModulii}, the values of $B_{0}$, $C_{44}$ and $C^\prime$ are shown as a function of the $Mn$ content, where 

\[ C^\prime= \frac{C_{11}-C_{12}}{2}\] 

\indent  is the elastic modulus related to $\langle 1 1 0\rangle \langle 1 \bar{1} 0 \rangle $ shear. All the alloys satisfy the Born stability criteria, $B_{0}>0$, $C_{44}>0$, and $C^\prime>0$. Experimental values  in other $Cu-Al$ based systems posses values in the same range than the ones calculated here. For instance, for $Cu-Al-Be$ with $Be$ contents between $2.5-5 at. \% $, the room temperature values are in the ranges $B_{0}$ $\approx$ $125-130$ $GPa$, $C_{44}$ $\approx$ $90-95$ $GPa$, and $C^\prime$ $\approx$ $7-9$ $GPa$ \cite{planes1996vibrational}. For $Cu-Al-Ni$, $B_{0}$ $\approx$ $127-133$  $GPa$, $C_{44}$ $\approx$ $95$ $GPa$ and $C^\prime$ $\approx$ $7.34-7.48$ $GPa$ \cite{manosa1994elastic}. In the compilation by Romero and Pelegrina \cite{romero2003change} for $Cu-Zn$, $Cu-Al-Zn$, $Cu-Al-Be$ and $Cu-Al-Ni$ alloys, $C_{44}$ ranges between around $95$ and $105$ $GPa$ at different temperatures. \\
\indent  As shown in Fig. \ref{fig:ElasticModulii}, $C^\prime$ has a low value as compared to the other two moduli in all the range of compositions and decreases with the amount of copper in the alloy. The comparatively low values of $C^\prime$ indicate that the bcc structure has weak restoring forces for shears in the $\lbrace 1 1 0 \rbrace$  planes along the directions $\langle 1 \bar{1} 0 \rangle $ \cite{planes2001vibrational} and play an important role in the occurrence of the martensitic transformation \cite{verlinden1984third}.

The Young and shear moduli, and the Poisson ratio, obtained by Eqs. \ref{GH} - \ref{nuIsot} are listed in Table \ref{tab:Relaciones}. For $Cu_{2}AlMn$ our results closely compares with those of Ref. \cite{wen2018first}.It should be noted that both the Young and shear moduli increases and the value of Poisson ratio decreases when the manganese content grows.

\begin{table*}
\begin{center}
\begin{tabular}{| c | c | c | c | c | c | c | c | c | c |}\hline Alloy  & $E[GPa] $ & $ G_H [GPa] $ & $ \nu$ & $ A^C $ & $ B_0/G  $ & $C_{12}/C_{44}$ & $\Theta_{D, 0}[K]$  \\ \hline
$ Cu_3 Al $ & $ 99.00 $ & $ 36.02 $ & $ 0.374 $ & $ 24.17 $ & $ 3.64 $ & $1.29$ & $322.14$ \\ \hline
$ Cu_{11}Al_4 Mn $ & $ 102.61 $ & $ 37.54 $ & $0.367 $ & $ 17.04 $ & $ 3.41 $ &   $1.26$ & 330.57 \\ \hline
$ Cu_{10}Al_4 Mn_2 $ & $ 106.63 $ & $ 39.31 $ & $ 0.356 $ & $ 11.27$ & $ 3.14 $ & $1.25$ & 340.56 \\ \hline
$ Cu_9 Al_4 Mn_3 $ & $ 112.00 $ & $ 41.59 $ & $ 0.346 $ & $ 10.53 $ & $ 2.93 $ & $1.17$ & 351.97\\ \hline
$ Cu_2 Al Mn $ & $ 126.12  $ & $ 47.57  $ & $ 0.325 $ & $ 7.44 $ & $ 2.53  $ & $1.09$ & 378.18 \\ \hline
$ Cu_2 Al Mn $\cite{wen2018first} & $ 123.9 $ & $ 46.6 $ & $0.331 $ & $ 7.68 $ & $ 2.63  $ & $1.11$ & 375.40\\ \hline
\end{tabular}
\caption{Calculated Young and shear moduli, Poisson ratio, anisotropy, ductility, Cauchy relations and Debye temperature at the equilibrium volume at $T=0K$}
\label{tab:Relaciones}
\end{center}
\end{table*}

\indent  The anisotropies of alloys crystals, $A^{C}$, calculated as the Zener ratio \cite{zener1947}:
\[ A^C = \frac{C_{44}}{C^\prime} \]
are also listed in Table \ref{tab:Relaciones}. The low values of $C^\prime$ lead to high values of the anisotropy. According to the compilation of experimental data by Z. Lethbridge et al. \cite{lethbridge2010elastic}, all cubic materials with $A^{C}>4$ show a negative Poissons ratio in some combination of load direction /  transversal plane. Materials with negative Poisson ratio are called auxetics \cite{evans1991ij}, and have the property of displaying a widening upon application of a longitudinal tensile strain. Around $69$  $\%$ of the elemental cubic metals posses negative Poisson ratio when stretched along the $[1 1 0]$ direction \cite{baughman1998negative}. In a recent work \cite{xu2020negative} it has been shown that an alloy of composition $Cu - 16.9 Al - 11.6 Mn$ (at. $\%$) posses a negative Poisson's ratio along the $[110]$ direction when the strain is measured along the transverse $[1 \bar{1}  0]$ direction: $\nu_{\langle [ 1 1 0], [1 \bar{1} 0] \rangle} = -0.51$, and a large and positive Poisson's ratio when the strain is measured along the $[0 0 1]$ transverse direction: $\nu_{\langle [1 1 0], [0 0 1]\rangle} = 1.34 $. Using the elastic constants obtained in the present calculations these Poisson's ratio can be calculated through the relations \cite{baughman1998negative}:

\begin{align*}
\nu_{\langle [ 1 1 0], [1 \bar{1} 0] \rangle}& =\\&\frac{-2C_{11}C_{44}+(C_{11}-C_{12})(C_{11}+2C_{12})}{2C_{11}C_{44}+(C_{11}-C_{12})(C_{11}+2C_{12})} \end{align*}

\begin{align*}
\nu_{\langle [1 1 0], [0 0 1]\rangle}& =\\&\frac{4C_{12}C_{44}}{2C_{11}C_{44}+(C_{11}-C_{12})(C_{11}+2C_{12})}
\end{align*}

 The obtained values are detailed in Table \ref{tab:PoissonRatios}. The present values are in the range of the experimental data.

\begin{table}
\begin{center}
\begin{tabular}{ | c | c | c | }  \hline
 Comp. & $\nu_{\langle [ 1 1 0], [1 \bar{1} 0] \rangle}$ & $\nu_{\langle [1 1 0], [0 0 1]\rangle}$ \\  \hline
$ Cu_{3}Al $ & $ -0.79 $ & $ 1.67 $ \\ \hline
$ Cu_{11}Al_{4} Mn $ & $ -0.72 $ & $ 1.57 $ \\ \hline
$ Cu_{10}Al_{4}Mn_{2}$& $ -0.61 $ & $ 1.41 $ \\ \hline
$ Cu_{9}Al_{4}Mn_{3} $ & $ -0.59 $ & $ 1.37 $  \\ \hline
$ Cu_2Al Mn $& $ -0.48 $ & $ 1.19 $ \\ \hline
\end{tabular}
\caption{Poisson's ratio for loads applied in the $[110]$ direction and strains along the transverse directions $[1 \bar{1} 0]$ and $[0 0 1]$.}
\label{tab:PoissonRatios}
\end{center}
\end{table}

\indent  The anisotropy is also related to the spinodal decomposition process. For $A^C>1$ , spinodal decomposition gives rise to compositional plane waves on $\{1 0 0\}$ planes, whereas for $A_C<1$ this occurs on $\{1 1 1\}$ planes  
\cite{cahn1961spinodal,cahn1962spinodal}. For the alloys studied in this work, the prediction of $A^C>1$ agrees with the experimental observation of composition modulations along the $ \langle 1 0 0 \rangle$ direction \cite{bouchard1975phase}.

\begin{table*}
\begin{center}
\begin{tabular}{ | c | c| c | c | c  |}  \hline 
 Alloy & Ref. & $ v_l[m/s] $ & $ v_t[m/s] $ & $ v_{eff}[m/s]$ \\  \hline
$ Cu_3 Al $ & This work & $ 5002.1 $ & $ 2242.2 $ & $ 2529.3 $  \\ \hline
$ Cu_{11}Al_4 Mn $ & This work & $5029.2 $ & $2308.7 $ & $2601.5 $    \\ \hline
$ Cu_{10}Al_4 Mn_2 $ & This work & $5070.3 $ & $ 2396.7 $ & $ 2696.9 $     \\ \hline
$ Cu_9 Al_4 Mn_3 $ & This work & $ 5129.8 $ & $ 2485.8 $ & $ 2793.5$    \\ \hline
 $ Cu_2 Al Mn $ & This work & $ 5290.6  $ & $ 2690.6  $ & $ 3015.2 $  \\ 
& \cite{wen2018first}, PAW & $ 5272.9 $ & $ 2650.4 $ & $ 2972.3 $  \\ \hline
\end{tabular}
\caption{Calculated isotropic longitudinal, transversal and average sound velocities.}
\label{tab:VAverage}
\end{center}
\end{table*}

\begin{table*}[t!]
\begin{center}
\begin{tabular}{ | c | c | c | c | c | c | c | c |}  \hline 
 Alloy & Ref. & $ v_l\langle100 \rangle $ & $ v_t\langle100 \rangle $ & $ v_l\langle110 \rangle$ & $ v_t\langle110 \rangle$ & $v_l\langle 111 \rangle$ & $ v_t\langle 111 \rangle$ \\  \hline
$ Cu_3 Al $ & This work & $ 4376.6 $ & $ 3718.9 $ & $ 5687.1 $  &$ 1133.2 $ & $ 6061.3 $ & $2244.6$   \\ \hline
$ Cu_{11}Al_4 Mn $ & This work & $4393.4 $ & $3749.4 $ & $5704.0 $  & $1284.4 $ & $  6078.4$  & $ 2288.2 $  \\ \hline
$ Cu_{10}Al_4 Mn_2 $ & This work & $4436.7 $ & $ 3717.5 $ & $ 5681.3 $ & $ 1566.2 $ & $ 6039.4 $  & $ 2398.0 $  \\ \hline
$ Cu_9 Al_4 Mn_3 $ & This work & $ 4463.9 $ & $ 3822.4 $ & $ 5757.5 $ & $ 1665.9$ & $ 6128.4 $ & $ 2407.3 $   \\ \hline
& This work & $ 4595.7  $ & $ 3939.8 $ & $ 5878.4 $ & $ 2042.7 $ & $   6247.7$  & $ 2562.2 $ \\ 
$ Cu_2 Al Mn$ & \cite{wen2018first}, PAW & $4576.2 $ & $ 3942.1 $ & $ 5882.5 $ &  $1938.2$  & $ 6257.7 $ & $ 2536.1 $ \\ 
& \cite{michelutti1978magnetization}, Exp & $4553 $ & $ 3787 $ & $ 5670 $ & $---$ & $ 6003 $ & $ --- $ \\ \hline
\end{tabular}
\caption{Calculated anisotropic sound velocities. }
\label{tab:DirectionalSoundVelocities}
\end{center}
\end{table*}

\indent  The quotient between the bulk modulus and the shear modulus gives an idea about the ductility of the material. If this quotient is greater than 1.75, it is said that the material is ductile \cite{pgh1954xcii}. According to this classification and the results shown in Table \ref{tab:Relaciones}, all the studied alloys are predicted to show a ductile character.  The Cauchy relation among $C_{12}$ and $C_{44}$ is also given in Table \ref{tab:Relaciones}. This ratio indicates whether the interatomic forces are central or noncentral: if the quotient is equal to $1$ the forces are central \cite{pettifor1992theoretical}.\\
\indent  The calculated Debye temperatures at the equilibrium volume, $\Theta_{D,0}=\Theta_{D}(V_{0})$ are shown in the last column of Table \ref{tab:Relaciones}. It can be seen that the Debye temperature increases with the $Mn$ content; this is consistent with the corresponding decrease of the Poisson ratio. For $Cu_{2}AlMn$, our result ($378.18$ $K$) can be compared with the experimental values found in the literature: Fenander et al. \cite{fenander1968low} obtained the value of $330 K$ using calorimetric techniques, whereas in Ref. \cite{michelutti1978magnetization} a value of $372 K$ was obtained by means of neutron scattering measurements. From the theoretical side, a value of $375.40 K$ was obtained by Wen et al. \cite{wen2018first}.\\ 
\indent  From the elastic constants it is also possible to calculate the isotropic longitudinal, $v_{l}$, and transverse, $v_{t}$, components of the sound velocity as:

\begin{equation}
\rho v_l^2 = B_{0} + \frac{4}{3} G_{H}
\end{equation}

\begin{equation}
\rho v_t^2 = G_{H}
\end{equation} 

and its effective or average value

\begin{equation}
v_{eff} = \left[\frac{1}{3}\left(\frac{1}{v_l^3} + \frac{2}{_t^3} \right) \right]^{-1/3}
\end{equation}

 The obtained results are given in Table \ref{tab:VAverage}. For $Cu_{2}AlMn$ our results compare well with calculations presented in \cite{wen2018first}.\\
\indent  The velocities of longitudinal and transverse elastic waves in the direction $\langle 100 \rangle $, $\langle 110 \rangle$ and $ \langle 111 \rangle$ for cubic crystals can be calculated by means of the second order elastic constant and the mass density $\rho$, using the relations \cite{truell2013ultrasonic}:

\[ v_l\langle 100 \rangle = \sqrt{\frac{C_{11}}{\rho}}  \]
\[ v_t\langle 100 \rangle = \sqrt{\frac{C_{44}}{\rho}} = v_{t_1}\langle 110 \rangle \]
\[ v_l\langle 110 \rangle = \sqrt{\frac{C_{11}+C_{12} + 2C_{44}}{2 \rho}} \]

\[ v_{t_2}\langle 110 \rangle = \sqrt{\frac{C_{11}-C_{12}}{\rho}}\]
\[ v_l\langle 111 \rangle = \sqrt{\frac{C_{11}+2C_{12} +  4 C_{44}}{3 \rho}}\]
\[ v_t\langle 111 \rangle = \sqrt{\frac{C_{11}-C_{12} +  C_{44}}{3 \rho}}\]

\indent The obtained values, and a comparison with recent calculations by \cite{wen2018first} and experimental values from \cite{michelutti1978magnetization} are given in Table \ref{tab:DirectionalSoundVelocities}. As a general trend, both the average and the anisotropic sound velocities increase as the $Mn$ content does. In particular, the $v_t \langle 1 1 0 \rangle$ is sensibly lower for $ Cu_3Al $ than for $ Cu_2AlMn $, even though for the remaining directions the speeds do not differ  markedly. The values obtained in this work for the directional velocities in $Cu_{2}AlMn$ are in good agreement with the values calculated in \cite{wen2018first}, although slightly overestimate the experimental values from \cite{michelutti1978magnetization}.

\subsection{Thermal properties}
\label{sbc:QHA}

\begin{figure}[t!]
\centering
\includegraphics[width=0.45\textwidth]{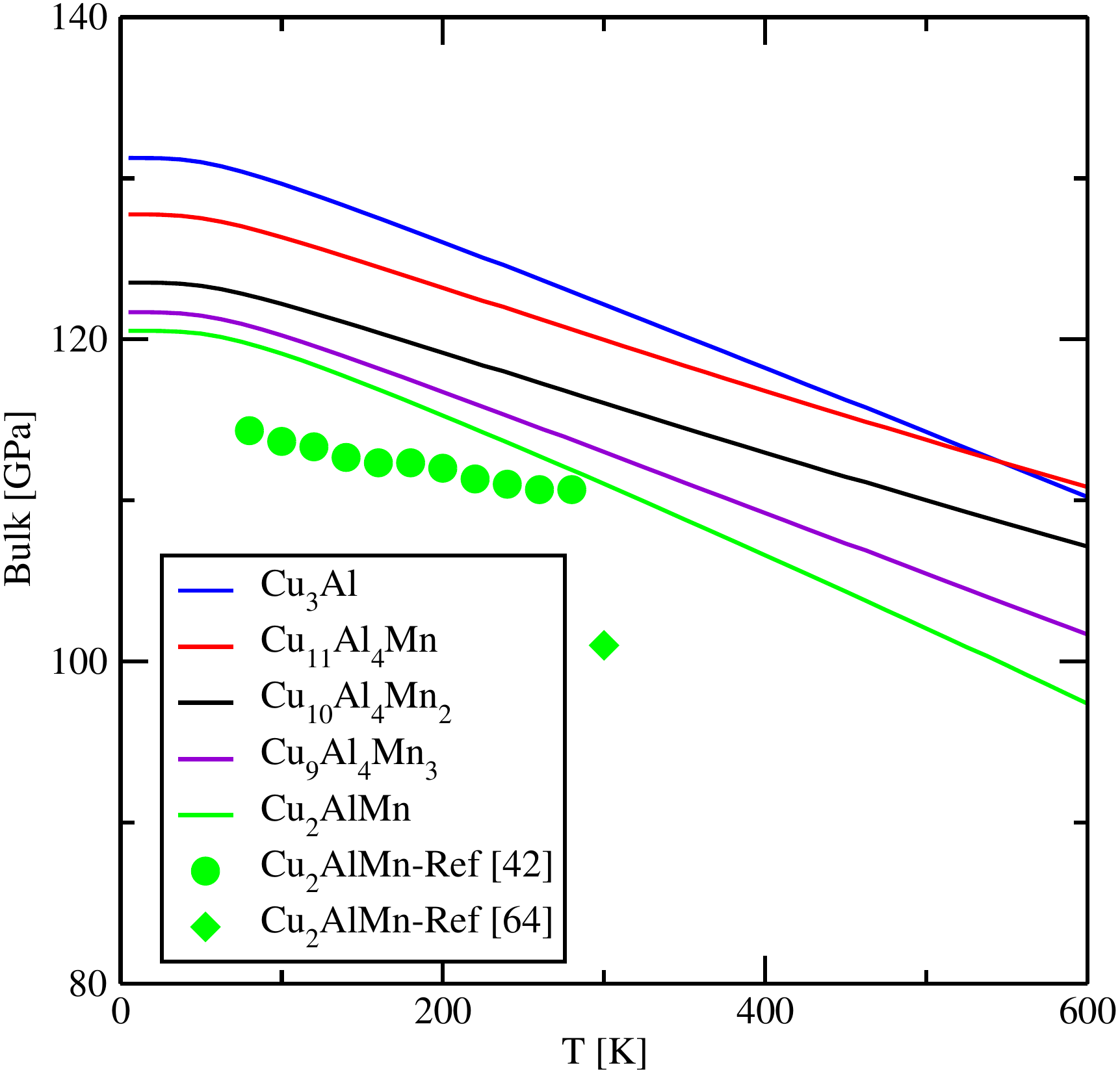}
\caption{Thermal variation of the bulk modulus for different compositions}
\label{fig:BulkvsT}
\end{figure}

\begin{figure}[t!]
\centering
\includegraphics[width=0.45\textwidth]{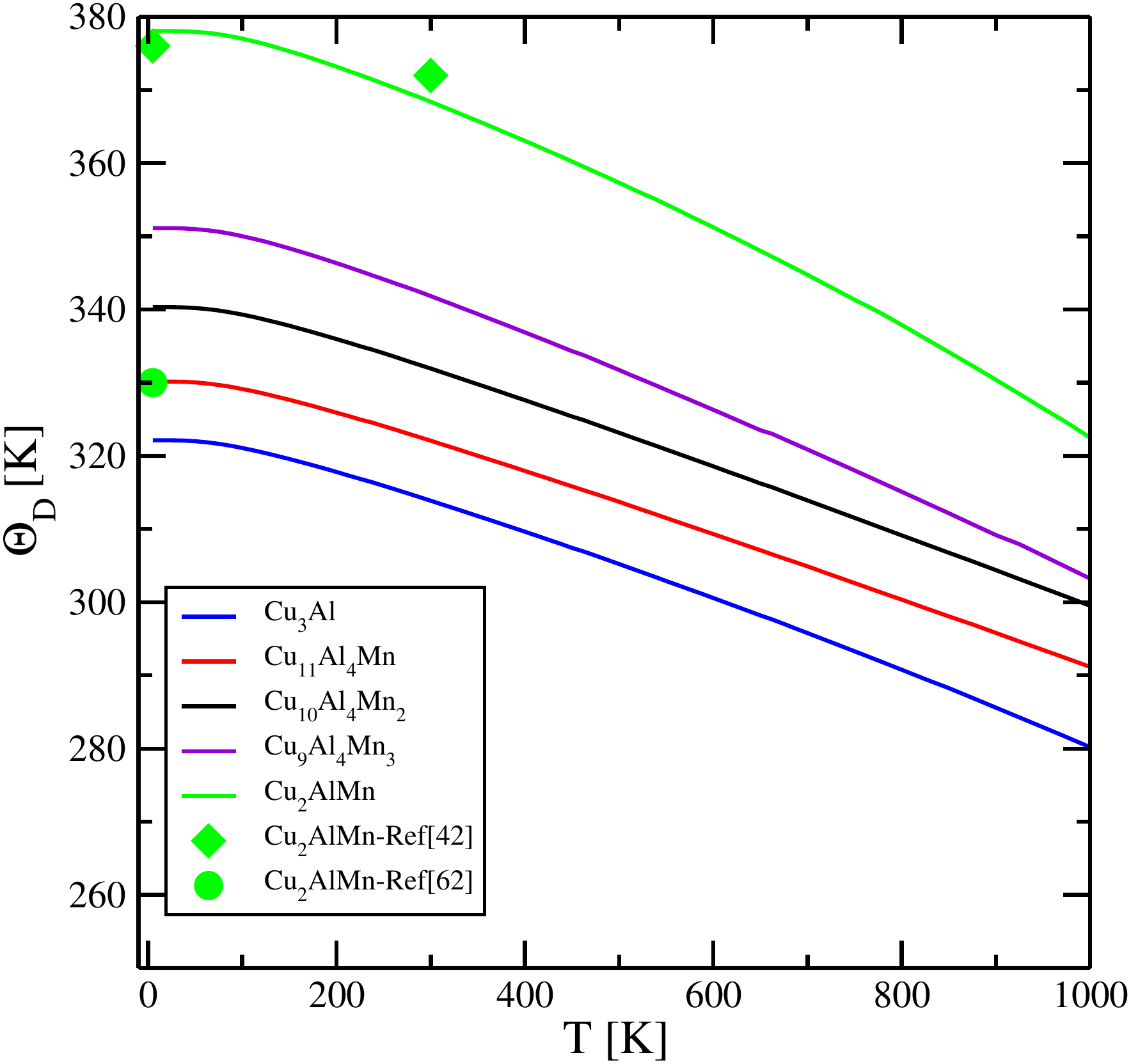}
\caption{Thermal variation of the Debye temperature for different compositions.  The experimental values for $Cu_2AlMn$ are represented with diamonds \cite{michelutti1978magnetization} and circles \cite{fenander1968low}.}
\label{fig:DebyevsT}
\end{figure}

\indent The thermal properties of the alloys, between $0$ and $600$ $K$, were studied by means of the quasi-harmonic approximation detailed in Subsection \ref{sb:Thermalproperties}. The variation of the isothermal bulk modulus in function of temperature for the different alloys is shown in Fig. \ref{fig:BulkvsT}. Experimental values for $Cu_{2}AlMn$ obtained from the literature are included for comparison \cite{michelutti1978magnetization,green1977plastic}. The calculated values overestimate the experimental ones from Ref. \cite{michelutti1978magnetization} by less than $5 \%$ at low temperatures, but the differences reduce as room temperature is approached.\\
\indent The thermal variation of the Debye temperature for the different alloys is show in Fig. \ref{fig:DebyevsT}. The experimental data from Ref. \cite{michelutti1978magnetization,fenander1968low} for $Cu_2AlMn$ are included for comparison; our results are in good agreement with the measurements by Michelluti et al \cite{michelutti1978magnetization} It can be seen that the $\Theta_D$ is almost invariable when $T < 100 K$. For greater temperatures than $T = 100$ the $\Theta_D$  decreases. 

\indent  Thermal expansion is an important phenomenon in the thermodynamics of materials.  It is interesting to analyze the behavior of the lattice parameter as a function of temperature. This is displayed in Fig. \ref{fig:latt300} for the different alloys considered in this work. The slopes of the lattice parameters are very similar for the different compositions; there are not remarkable difference between the predicted thermal variation of the lattice parameters for $Cu_3Al$ and for $Cu_2AlMn$. The values of the thermal expansion coefficient at room temperature are $5.79 \times 10^{-5}$ $K^{-1}$ for $Cu_3Al$ and  $6.06 \times 10^{-5}$ $K^{-1}$ for $Cu_2AlMn$. \\
\indent In Fig. \ref{fig:latt300} the lattice parameters at room temperature are plotted as a function of the alloy composition. The lattice parameter increases with the $Mn$ content in an approximately linear way. Also included in Fig. \ref{fig:latt300} are the experimental data from  Refs. \cite{kainuma1998phase,sugimoto1998giant}. Our results slightly overestimates the experimental values; the maximum difference is below $1 \%$ in the Cu-rich corner and gradually reduces for higher $Mn$ content. 

\begin{figure}[t!]
\centering
\includegraphics[width=0.45\textwidth]{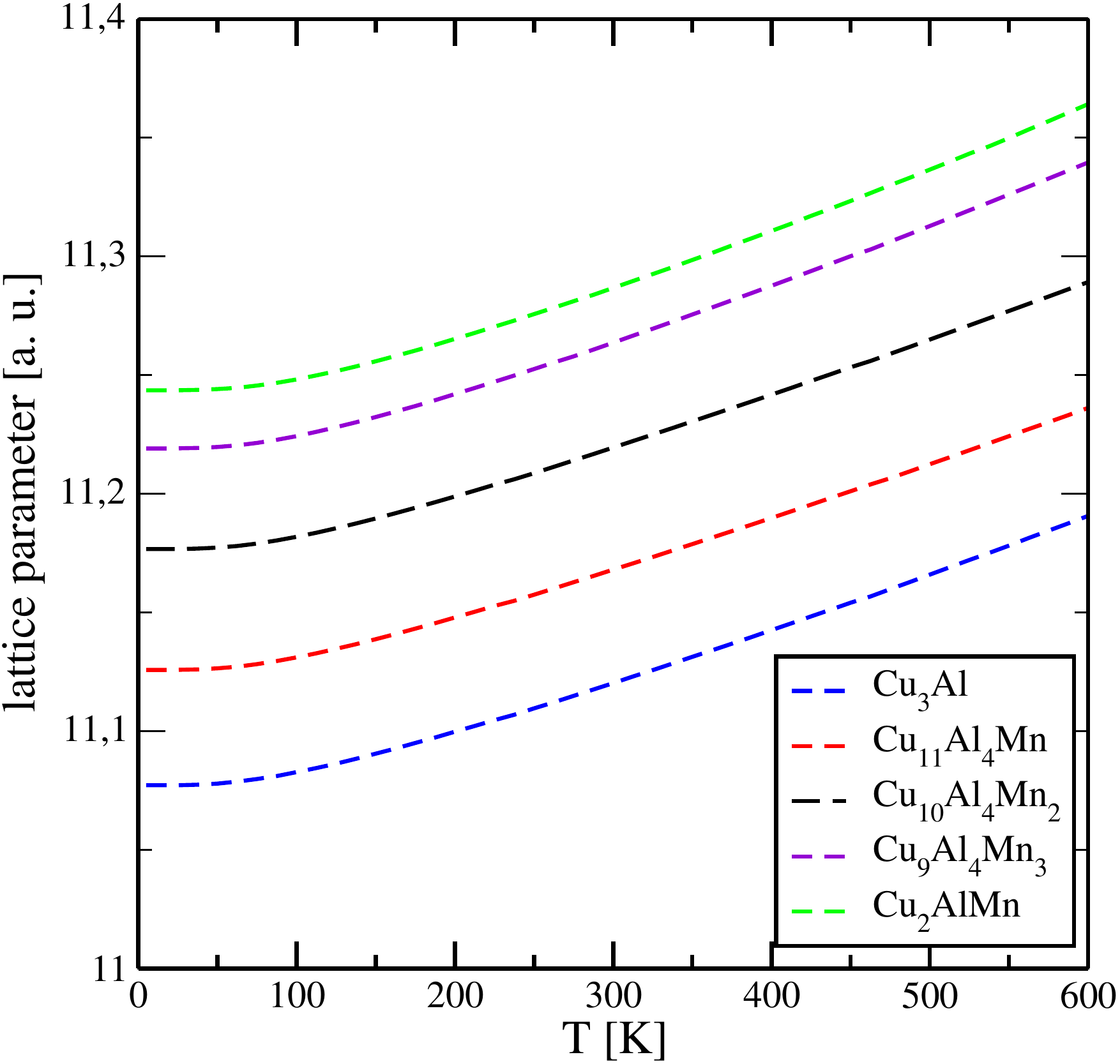}
\caption{Lattice parameter for the alloys at different temperatures.}
\label{fig:latt}
\end{figure}

\begin{figure}[t!]
\centering
\includegraphics[width=0.45\textwidth]{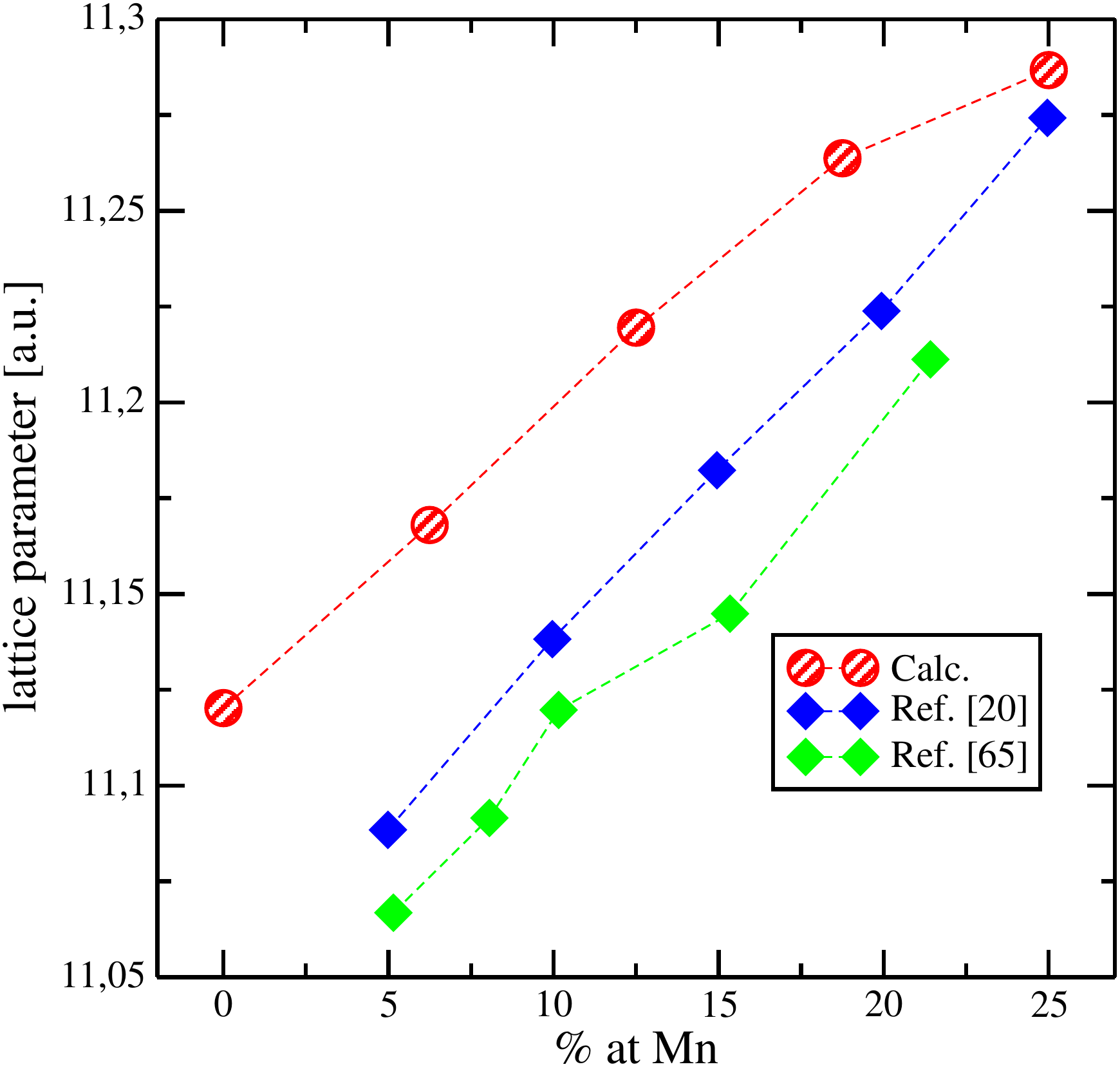}
\caption{Lattice parameter for the alloys at room temperature and comparison with experimental data}
\label{fig:latt300}
\end{figure}

\section{Conclusions}
\label{sc:Concl} 
\indent  In this work, the elastic constants of $Cu-Al-Mn$ alloys with structure derived from the bcc, and located along the pseudo-binary line $ Cu_3Al \to Cu_2AlMn $, have been determined by first-principles calculations. Our results present good agreement with both experimental and theoretical values reported in the bibliography. Although the equilibrium volume of the different alloys smoothly increases with the manganese content, the elastic constants $C_{11}$ and $C_{44}$ do not display remarkable composition dependence. A more noticeable change is found in the behavior of $C_{12}$, which steadily decreases with the $Mn$ content. For the alloys with lower $Mn$ content the constant $C^\prime$ is smaller, and the alloys become more susceptible to structural changes  under compression in direction $\langle 1 1 0\rangle$ \cite{verlinden1984third}. 
From the aforementioned elastic constants, other quantities of interest for the structural behavior of the alloy, such as Young's modulus, shear modulus and Poisson's ratio, have been obtained. \\
\indent The anisotropy has, in general, large values for all the studied alloys, being lower for $ Cu_2AlMn $ and increasing as the copper content grows. This would facilitate both spinodal decomposition and martensitic transformation\cite{verlinden1984third,cahn1961spinodal,cahn1962spinodal}. The directional Poisson's ratio for loads in the $[110]$ direction have been calculated. For strains along the transverse $[1 \bar{1} 0]$ direction a negative Poisson ratio is predicted, whereas for strains along $[1 0 0]$, the Poisson ratio is positive and greater than unity. These results agree with a recent experimental work \cite{xu2020negative}. \\
\indent The isotropic and directional sound velocities have been calculated from the elastic constants. Both values agree well with the known results for $ L2_1-Cu_2AlMn $. \\
\indent Finally, the behaviour of the bulk modulus, Debye temperature and equilibrium lattice parameter as a function of temperature has been calculated through the Quasi-harmonic approximation.  According with our calculations, the difference between the lattice parameter of $Cu_3Al$ and $Cu_2AlMn$ does not change significantly with the temperature. The agreement with experimental results has been found to be satisfactory\cite{michelutti1978magnetization,fenander1968low,green1977plastic}.

\section*{Acknowledgment}
The authors thank to UNCPBA, ANPCyT (PICT 2017–4062) and CONICET (Argentina) for their financial support, and to Dr. R. Romero (IFIMAT-UNCPBA) for helpful suggestions. A.A. thanks the post-doctoral fellowship from CONICET.

\printbibliography

\end{document}